\documentclass[twocolumn,aps,prl]{revtex4}
\usepackage{psfig}
\usepackage{epsfig}

\begin{document}
\title{Lengthscales and Cooperativity in DNA Bubble Formation}
\author{Z. Rapti$^{1}$, A. Smerzi$^{2,3}$,  K. \O. Rasmussen$^{2}$ and A. R. Bishop$^{2}$ }
\affiliation{$^1$ Center for Nonlinear Studies, Los Alamos National Laboratory, Los Alamos, New Mexico 87545, and School
of Mathematics, Institute for Advanced Study, Einstein Drive 1, Princeton, New Jersey 08540}
\affiliation{$^2$ Theoretical Division and Center for Nonlinear Studies,
Los Alamos National Laboratory, Los Alamos, New Mexico 87545}
\affiliation{$^3$ Istituto Nazionale di Fisica per la Materia BEC-CRS,
Universit\`a di Trento, I-38050 Povo, Italy} 
\author{C.H. Choi, and A. Usheva }
\affiliation{Beth Israel Deaconess Medical Center and Harvard Medical School,
99 Brookline Avenue, Boston, Massachusetts 02215}

\date{\today}

\begin{abstract}
It appears that thermally activated DNA bubbles of different sizes play central roles in important 
genetic processes. Here we show that the probability for the formation of such bubbles 
is regulated by the number of soft AT pairs in specific regions with lengths which at physiological temperatures
are of the order of (but not equal to) the size of the bubble. 
The analysis is based on the Peyrard-Bishop-Dauxois model, whose
equilibrium statistical properties have been
accurately calculated here with a transfer integral approach.
\end{abstract}
 
\maketitle

The genetic code underlying all forms of life is encoded in the DNA molecule by the four bases
guanine (G), thymine (T), adenine (A), and cytosine (C) strung along a sugar-phosphate backbone 
in a particular sequence. The four bases are, through hydrogen bonding, pairwise complementary 
(A-T and G-C) allowing the coding strand and it's complement to form the characteristic 
double helical DNA macromolecule. Although this construct is extraordinarily stable, it is clearly necessary 
that the double strands be separated in biological processes, including gene transcription, where the 
code is read by the appropriate protein machinery in the cell. It has long been an experimental fact \cite{stjaalet} that 
the DNA double-strand can be thermally destabilized locally to form temporary single stranded ``bubbles'' in the 
molecule. This local melting is made possible by the entropy gained by transitioning from the 
very rigid
double-strand to the much more flexible single-strand, which already at biologically relevant temperatures can 
balance the energy cost of breaking a few base pairs. Considering this entropic effect together with the 
inherent energetic heterogeneity --
GC base pairs are 25 \% more strongly bound than the AT bases --
of a DNA sequence, it is conceivable that certain regions (subsequences) are more 
prone to such thermal destabilization than others: This has been confirmed by model 
calculations as well as 
experiments. We have previously argued \cite{donald} that such regions may indeed experimentally
coincide with transcription initiation and regulatory sites. In this
way, the DNA molecule may help initiate its own 
transcription by containing bubble forming subsequences at the crucial positions in the sequence
where the transcription machinery assembles and engages its operation. 
If a robust general link between 
the formation of large thermal bubbles and transcription initiation is
sufficiently established, it becomes
crucially important to be able to accurately predict the subsequence of DNA
with propensity for the formation of bubbles of appropriate sizes. 

Here we show that the probabilities of finding bubbles extending over $n$ sites 
do not depend on a specific DNA subsequences.
Rather, such probabilities depend on the density of soft A/T base pairs within specific regions of length $L(\kappa)$. This
characteristic length is of the order of the size $n$ of the bubble at physiological temperatures,
but it diverges as the DNA melting temperature is approached. Our results
are based on a calculation of the thermal equilibrium statistical properties of the Peyrard-Bishop-Dauxois (PBD) 
model \cite{pb,dpb} using a transfer integral operator (TIO) technique. This model constitutes a very powerful tool to 
not only predict bubble formation probability in a given sequence but also to understand the underlying physical 
mechanisms \cite{nonlinearity}. Our previous  study of the PBD model
has been performed using Langevin \cite{donald,george} and Monte Carlo techniques \cite{kim}.  However, since our interest is 
centered on a very small portion of the
thermodynamical equilibrium state, namely on the formation of large bubbles, 
dynamical and iterative samplings as offered by these methods are not very efficient. 
Therefore, we have developed here a 
semi-analytic approach based on the TIO \cite{scalp,zhang} that allows
us to efficiently calculate 
relevant thermodynamical probabilities. 

The potential energy of the PBD model, in its simplest form, reads
\begin{equation}
{E}=\sum_{\kappa=1}^{N} \left[
V(y_n)+W(y_n,y_{n-1})\right] = \sum_{\kappa=1}^N {\cal E}(y_n,y_{n-1}),
\label{ham}
\end{equation}
where $V(y_n)=D_n (e^{-a_n y_n}-1)^2$, represents the nonlinear hydrogen bonds between the bases. 
$W(y_n,y_{n-1})=\frac{k}{2}\left(1+\rho e^{-b(y_n+y_{n-1})}\right)(y_n-y_{n-1})^2$ is the nearest-neighbor 
coupling that represents the (nonlinear) stacking interaction
between adjacent base pairs: it is comprised of a harmonic coupling with a state depended coupling constant 
effectively modeling the change in stiffness as the double strand is opened (i.e. entropic effects).
This nonlinear coupling results in long-range cooperative effects, leading to a sharp entropy-driven denaturation 
transition \cite{dpb,dp}. The sum in Eq.(\ref{ham}) is over all base-pairs of the molecule and $y_n$ denotes the relative
displacement from equilibrium bases at the $n^{th}$ base pair. The importance of the heterogeneity of the sequence is 
incorporated by assigning different values to the parameters of the Morse potential, depending on the the base-pair type.
The parameter values we have used are those from Refs. \cite{campa, para} chosen to reproduce a variety of thermodynamic 
properties.

{\it Transfer Integral Method.} 
All equilibrium, thermodynamic properties of the model (\ref{ham}) can be obtained through the partition function
\begin{eqnarray}
{\cal Z} &=&\int \prod_{\kappa=1}^N dy_n e^{-\beta{\cal E}(y_n,y_{n-1})} \label{parf}\cr 
&=&\int \prod_{n=s}^{s+\kappa-1} dy_nZ_\kappa(s)\,e^{-\beta{\cal E}(y_n,y_{n-1})},
\end{eqnarray}
where the notation
$$Z_{\kappa}(s)=\int \prod_{n \ne s,...,s+\kappa-1}^N  dy_n\, e^{-\beta {\cal E}(y_n,y_{n-1})}$$ has been introduced. 
$\beta= (k_BT)^{-1}$ is the Boltzmann factor.
In order to evaluate the partition function (\ref{parf}) using the TIO method, we first symmetrize 
$e^{-\beta {\cal E}(x,y)}$ by introducing \cite{dp} 
\begin{eqnarray}
S(x,y)&=&\exp\left(-\frac{\beta}{2} (V(x)+V(y)+2 W(x,y))\right) \label{kernel}\cr
      &=&S(y,x). \nonumber
\label{symker}
\end{eqnarray}
Here the second equality holds only when $x$ and $y$ correspond to base-pairs of the same kind.
Using Eq. (\ref{parf}) the expression for $Z_\kappa(s)$ is rewritten as 
\begin{eqnarray}
Z_\kappa(s)=&&\int \left(\displaystyle \prod_{n \ne s,...,s+\kappa-1}^N dy_n S(y_n,y_{n-1}) \right)\cr
&&\times dy_0 e^{ -\frac{\beta}{2} V(y_1)} e^{ -\frac{\beta}{2} V(y_N)},
\label{sparf}
\end{eqnarray}
where open boundary conditions at $n=1$, and $n=N$ have been used.
To proceed, a Fredholm integral equations with a real symmetric kernel
\begin{eqnarray}
\int dy S(x,y) \phi(y)=\lambda \phi(x)
\label{intsym}
\end{eqnarray}
must be solved separately for the A/T and for the G/C base-pairs. 

Since the eigenvalues are orthonormal and the eigenfunctions form a complete basis, Eq.(\ref{intsym}) can be used 
sequentially to replace all integrals by matrix multiplications in Eq. (\ref{sparf}). 
Whenever the sequence heterogeneity results in a non-symmetric $S(x,y)$, Eq.(\ref{intsym}) cannot be
used and we resort to a symmetrization technique, based on successive introduction of auxiliary integration variables, 
as explained in Ref. \cite{thesis}. 

As noted, in order to quantify the sequence dependence on DNA's ability to form bubbles of different sizes, 
we have previously monitored the frequency of opening events using Langevin and Monte Carlo
simulation techniques. Since the large openings constitute relatively rare events 
such techniques are not efficient (although essential for evaluating dynamical and non-equilibrium properties).
It is much more effective to imply the probabilities of large bubbles at a given site in the sequence 
directly from the thermodynamic distributions using the TIO. Importantly, we have confirmed below that this equilibrium 
approach reproduces the bubble locations observed by Langevin simulations for the same sequences. 
This suggests that the bubbles -- although large bubbles are rare events -- are governed by equilibrium statistics.

We evaluate the probabilities $P_{\kappa}(s)$,
for a base-pair opening spanning $\kappa$ base-pairs (our operational
definition of a bubble of size $\kappa$), starting at base-pair $s$ as
\begin{equation}
P_{\kappa}(s)={\cal Z}^{-1}{\int_t^{\infty}\prod_{n=s}^{s+\kappa-1} dy_n Z_\kappa(s)e^{-\beta{\cal E}(y_n,y_{n-1})}},
\label{prob}
\end{equation}
where $t$ is the separation (which we have taken as 1.5 {\AA}) 
of the double strand above which we define the strand to be melted.
 
{\it Numerical Results.} 
Using this technique, we are able to systematically investigate the relation between a given sequence containing
a (apparently) disordered mixture of A/T and G/C pairs and the probability of spontaneous, thermally activated, 
bubbles of 
various sizes. Our analysis
begins with a thorough study of  
two viral promoter sequences, Adeno major late promoter (AMLP) and 
Adeno Associate viral promoter (AAV P5). We have previously investigated 
the dependence of the thermally induced large bubbles in 
these sequences \cite{george,donald} and found that the opening profiles obtained through Langevin simulations 
of the PBD model agreed remarkably well with the 
local denaturation profiles indicated by S1 nuclease  experiments (see Ref. \cite{donald} for details). 
\begin{figure}[h]
\centerline{\psfig{file=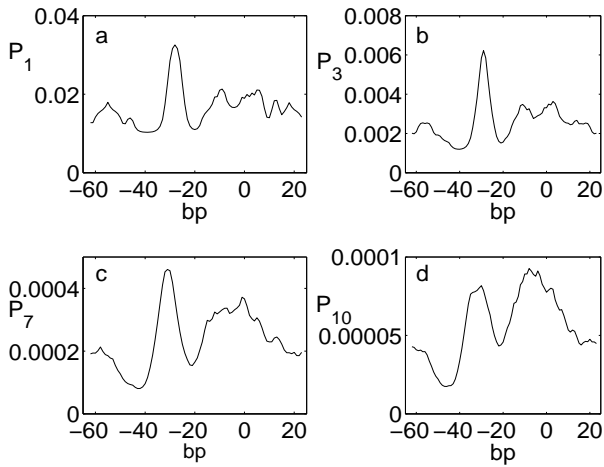,width=80mm,height=65mm}}
\caption{Probabilities $P_\kappa$ of creating 
bubbles spanning $\kappa=1,3,7$, and 10-bp, respectively, for the Adeno major late promoter at T=300K.}
\vspace{-.3cm}
\label{admlp}
\end{figure}
Here we use the TIO to calculate the probabilities Eq. (\ref{prob}) for the thermal creation of bubbles of size 1,3,7, and 10
base-pairs  for the AdMLP promoter at $T=300~K$ (Fig. \ref{admlp}).
The significant feature of the sequence is the occurrence of a TATA-box at base-pair location $-30$ 
with 7 consecutive A/T base-pairs. Around +1 there is rich region containing
$\sim 12$ A/T base pairs, which, however, 
are not located consecutively since a comparable amount of G/C
pairs are alternately embedded among the A/T pairs. 
Since A/T base-pairs are more weakly bound (softer) than G/C pairs, we could reasonably expect that
bubbles have a predominant  opening probability in the region -30. This is indeed the case for small bubbles, 
Figs. \ref{admlp}a and \ref{admlp}b .  
However, surprisingly, this prediction breaks down when considering bubbles of larger sizes.
The corresponding probability increases around bp $+1$ (Fig. \ref{admlp}c), up to the point that, for a bubble of size 
10 bp, it becomes the highest (Fig. \ref{admlp}d). 
This finding illustrates the strong interplay  between
the sequence of base-pairs and the size of the bubble in the thermal activity of DNA 
(Indeed, it is likely that bubbles of different sizes may initiate different genetic processes).
\begin{figure}[h]
\centerline{\epsfig{file=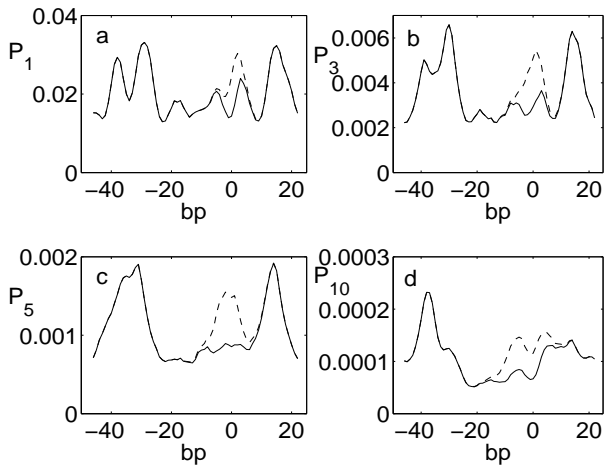,width=80mm,height=65mm}}
\caption{Probabilities $P_\kappa$ of creating bubbles of 1 bp ($\kappa=1$) length (a), 3 bp (b), 5 bp (c) and  10 bp (d) length, 
for the wild (dashed line) and mutant (solid line) P5 promoter.}
\vspace{-.5cm}
\label{p5}
\end{figure}

The replacement of 1 or 2 soft A/T with hard G/C base pairs in specific regions of the DNA can also hugely affect the 
probability for the formation of bubbles of given sizes. This is
illustrated with the AAV P5 promoter (Fig.\ref{p5}). 
This sequence regulates the AAV gene expression, and it has been shown \cite{Usheva} to bind the transcription 
initiator Yin Yang 1 (YY1) and to be active for TATA-Box protein (TBP)-independent transcription. The mutation of this 
promoter in which the two A/T bases at $+1$ and $+2$ are  replaced by two G/C 
bases, is known to destroy the binding site for the YY1 initiator and thereby inhibit transcription. 
We have previously shown by Langevin simulations of the PBD model
that this mutation also suppresses the formation of large bubbles around bp +1. 
Here we again calculate the probability to obtain bubbles of various
sizes using the TIO. In Fig. \ref{p5}  we show the probability of obtaining bubbles
of sizes n=1,3,5, 
and 10 for the wild (dashed line) and the mutated (solid line) AAV P5 promoter. 
The mutation causes a dramatic change in the double strand's ability 
to form large bubbles at and around the mutated region. However, the $P_1$ and the $P_{10}$ probabilities are much 
less affected by the mutation. Notice for the wild type P5 AAV promoter,
the region around the TATA-box has the largest probability for forming large bubbles (panel d).
It is important to note that the AAV P5 promoter has four A/T rich regions: four consecutive A/T's around position -40: seven A/T's around position -30; five A/T's at the transcription start site +1; six A/T's around  +14,
and all these soft regions are clearly discernible in $P_1$ (panel a). 
\begin{figure}[h]
\centerline{\epsfig{file=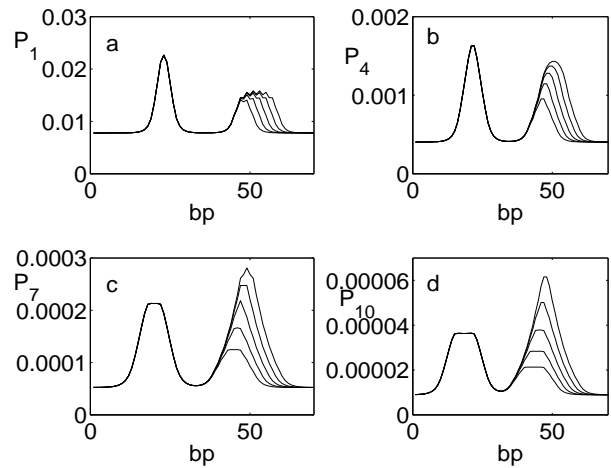,width=80mm,height=65mm}}
\begin{center}
\caption{Probability $P_\kappa$ for the formation of bubbles of sizes $\kappa=1,4,7,10$ bps.
The sequences are composed of 20 G/C, 
5 A/T, 20 G/C followed by different sequences comprising 3,4,5,6,7 A/T
alternating with G/C bps. The last 20 bps  are G/C. }
\label{inversion}
\end{center}
\end{figure}
From these results on the AAV P5 and AdMLP promoters we can speculate that the occurrence and intensity of a peak in the 
bubble probabilities does  not depend on the specific composition of
the DNA fragment. Rather, there is an essential 
interplay between the content of A/T and G/C base pairs and the size of the 
bubble being examined. Understanding the mechanisms regulating the DNA 
openings are of great importance for predicting and engineering DNA processes, and 
we therefore now consider a series of simple (but experimentally realizable) 
DNA sequences where the effects discussed above are reproduced in detail. 
Our purpose is to isolate the underlying mechanisms. 
Our five sequences are all composed of 20 G/C, 5 A/T and 20 G/C base pairs. This is followed 
by a sequence that alternates A/T and G/C base pairs. We use 3, 4, 5, 6, and 7 A/T base-pairs in the
five sequences. Finally, all five sequences are terminated with 20 G/C base-pairs.

As shown in  Fig. \ref{inversion}a, the largest 1 bp opening probability is localized at $+20$, 
a region that contains five consecutive A/T bases, and is therefore expected to be more susceptible to open than the region 
localized around $+50$, containing 
A/T's alternating with G/C's.
However, this simple picture changes dramatically as we move to
larger bubbles, Figs. \ref{inversion}b, \ref{inversion}c, and \ref{inversion}d.
In all these cases, the height of the second peak increases as compared to the peak at +20. With 3 and 
4 A/T's, the peaks saturate at a value lower than the first peak. However, 
the height of the two peaks for the sequence with 5 A/T's, becomes
equal in $P_7$, and remains so for larger bubbles. The sequence 
with 6 A/T's shows an inversion of the opening probability, similar to that
observed in the AdMLP sequence: at $P_{10}$ the most probable 10 base-pair opening 
occurs around the base-pair location $50$. These data indicate that the opening probability of a 
bubble of a given length
does not trivially depend on the number of consecutive A/T's in the DNA sub-sequence.
Instead, bubbles of sizes $n$ form with higher probabilities in regions where the 
number of A/Ts over some characteristic length $L(\kappa)$ is higher, even if the A/Ts are mixed with G/C pairs.

\begin{figure}[h]
\centerline{\psfig{file=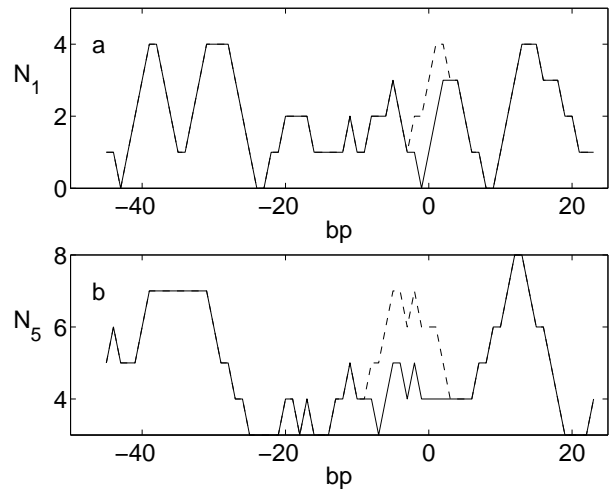,width=80mm,height=65mm}}
\caption{Number of A/T bps contained in the characteristic length $L(\kappa)$, where $\kappa=1$ (top panel), and $5$ 
(bottom panel) for the wild (dashed line) and mutant (solid line) P5 promoter.}
\label{number_at}
\end{figure}

To confirm this hypothesis we have extracted the characteristic lengths $L(\kappa)$ as a function of the bubble size $n$ 
from the probability distributions of the simple sequences
considered in Fig. \ref{inversion}. For instance, for $\kappa=1$, 
Fig. \ref{inversion}a , we have obtained 
$L(1)=4$ sites, 
while for $\kappa=5$, Fig. \ref{inversion}b, we have $L(5)=10$ sites. We have therefore considered 
the AAV P5 promoter DNA sequence of Fig. \ref{p5}. Starting from each site $s$ of the sequence,  
we count the number $N_\kappa(s)$ of 
A/T pairs contained over the corresponding next $L(\kappa)$ sites.
In Fig. \ref{number_at}a  we show the results for bubbles of size $\kappa=1$, which can be compared with
Fig. \ref{p5}a . The small difference between the mutant and the wild opening probability for the 
sites around located at $0$ is well reproduced. The difference between the wild and the mutant sequence 
is most pronounced for $\kappa=5$,
Fig. \ref{number_at}b, which is also in agreement with the TIO calculation shown  in Fig. \ref{p5}c.

\begin{figure}[h]
\centerline{\psfig{file=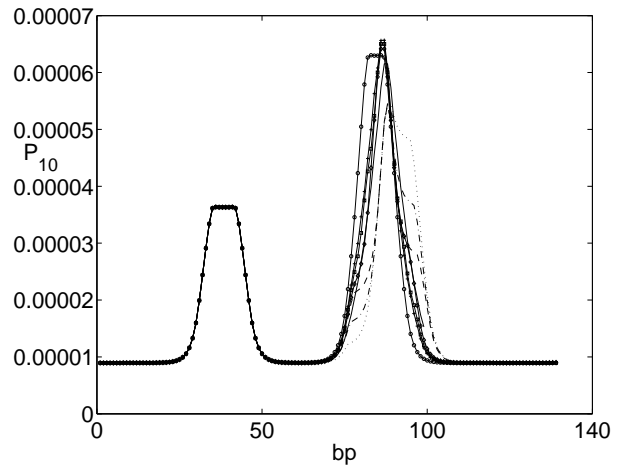,width=80mm,height=65mm}}
\caption{Probability $P_{10}$ for the formation of a bubble of 10 consecutive bps.
The sequences consist of 40 G/C, 5 A/T and 20 G/C followed by 14 bps 
containing different random combinations of 7 A/T and 7 G/C.
All sequences end with 40 G/C bps.}
\label{p_57}
\end{figure}
We have also demonstrated that the opening probability does not depend on the specific distribution of the 
AT pairs contained in the characteristic regions of length $L(\kappa)$. This is shown in  Fig.\ref{p_57}, where 
we have calculated with the TIO the 
probability for formation of bubbles with size $P_{10}$ for a sequence where the second 
A/T rich regions always contains 7 A/T distributed in several different combinations, but 
always over a 20-base region. We see that independently of the distribution of 
A/T base-pairs, the probability of 10 base-pair bubbles is always
largest in the right-most region. 
Physically, we interpret this
as the nonlinear coherence dominating (smoothing out the effect of) the base-pair disorder. 

In summary, we have developed a semi-analytical techniques (TIO)
that allows the efficient prediction of a given sequence 
for thermally induced bubbles of given 
sizes. We have found that large thermally induced bubbles arise 
through a subtle interplay between length scales inherent in the nonlinear dynamics, and the sequence disorder. 
Our results provide new understandings that can help to not only identify new protein coding genes, 
but also enable reverse-engineering for use in future gene therapeutic applications. 

This work at Los Alamos National Laboratory is supported by the US Department of Energy (contract No. W-7405-ENG-36)
and by a NIH grant for A.U. (Grant Number:  R01 GM073911).



\begin{thebibliography}{10}



\bibitem{stjaalet}M. Gueron, M. Kochoyan, and J.L. Leroy 
Nature {\bf 328},  89 (1987);  M. Frank-Kamenetskii Nature {\bf 328} 17 (1987).
\bibitem{donald} C.H. Choi, G. Kalosakas, K.\O. Rasmussen, M. Hiromura,
A. R. Bishop and A. Usheva, Nucleic Acids Res. {\bf 32}, 1584, (2004).
\bibitem{dpb} T.\ Dauxois, M.\ Peyrard and A.\ R.\ Bishop, Phys.\ Rev.\ E
{\bf 47} R44 (1993).
\bibitem{pb} M. Peyrard and A. R. Bishop, Phys. Rev. Lett, {\bf 62}, 2755, (1989).
\bibitem{nonlinearity} M. Peyrard, Nonlinearity {\bf 17}, R1 (2004).
\bibitem{george} G. Kalosakas, K.\O. Rasmussen, A. R. Bishop, C.H. Choi,
and A. Usheva, Europhys. Lett. {\bf 68}, 127, (2004).
\bibitem{kim} S. Ares, N.K. Voulgarakis, K.\O. Rasmussen and A.R. Bishop, 
Phys. Rev. Lett., {\bf 94}, 035504, (2005).
\bibitem{scalp} D. J. Scalapino, M. Sears and R.A. Ferrell, 
Phys. Rev. B, {\bf 6}, 3409, (1972).  
\bibitem{zhang} Y. Zhang, W.-M. Zheng, J.-X. Liu, and Y. Z. Chen, Phys. Rev. E, 
{\bf 56}, 7100, (1997).
\bibitem{dp} T.\ Dauxois and M.\ Peyrard, Phys.\ Rev.\ E {\bf 51} 4027 (1995).
\bibitem{campa} A. Campa and A. Giansanti, Phys. Rev. E, {\bf 58}, 3585, (1998).
\bibitem{para}The parameters were chosen in Ref. \cite{campa} to fit thermodynamic properties of DNA: 
$k=0.025eV/A^2$, $\rho=2$, $\beta=0.35A^{-1}$ for the inter-site coupling;
for the Morse potential $D_{GC}=0.075eV$, $a_{GC}=6.9A^{-1}$ for a G-C
bp, $D_{AT}=0.05eV$, $a_{AT}=4.2A^{-1}$ for the A-T bp.
\bibitem{thesis} M. B. Fogel, {\it Nonlinear Order Parameter Fields: I. Soliton Dynamics,
II. Thermodynamics of a Model Impure System}, Ph.D. Thesis, Cornel University, (1977).
\bibitem{Usheva} A. Usheva and Shenk, Cell {\bf 76}, 1115 (1994)






\end{thebibliography}
\end{document}